\documentclass[prl,aps,nofootinbib,twocolumn,showpacs]{revtex4}
\usepackage{amsmath}
\usepackage{amssymb}
\usepackage{amsthm}
\usepackage{amsfonts}
\usepackage{enumerate}
\usepackage{latexsym}
\usepackage[dvips]{graphicx}

\begin{document}

\title{On the sign structure of doped Mott insulators}
\author{ K. Wu,$^{1}$ Z.Y. Weng,$^{1}$ and J. Zaanen$^{2}$}
\affiliation{$^{1}$Center for Advanced Study, Tsinghua University, Beijing, 100084, China%
\\
$^{2}$Instituut Lorentz for Theoretical Physics, Leiden University, Leiden,
The Netherlands}

\begin{abstract}
We demonstrate that the sign structure of the t-J model on a hypercubic
lattice is entirely different from that of a Fermi gas, by inspecting the
high temperature expansion of the partition function up to all orders, as
well as the multi-hole propagator of the half-filled state and the
perturbative expansion of the ground state energy. We show that while the
fermion signs can be completely gauged away by a Marshall sign
transformation at half-filling, the bulk of the signs can be also gauged
away in a doped case, leaving behind a rarified \textquotedblleft
irreducible\textquotedblright\ sign structure that can be enumerated easily
by counting exchanges of holes with themselves and spins on their real space
paths. Such a sparse sign structure implies a mutual statistics for the
quantum states of the doped Mott insulator.
\end{abstract}

\date{\today }
\pacs{71.10.Fd,71.27.+a,02.70.Ss,74.20.Mn}
\maketitle

\tolerance 10000

The progress in the understanding of the physics of strongly interacting
electron systems has been strongly hindered by the infamous fermion minus
sign problem rendering field theoretical and statistical physics methods to
be ill behaved for fermions. The t-J model, catching the essence of the
doped Mott insulators, is archetypical. Despite twenty years of concerted
effort, inspired by its relevance towards the problem of high $T_{c}$
superconductivity,\cite{anderson} nothing is known rigorously about this
model, except then for the one dimensional case. In fact, the other
exception is the Mott-insulating state at half-filling, where the Hubbard
projection turns the indistinguishable fermions into distinguishable spins,
and the remnant signs of the unfrustrated spin problem can be gauged away by
a Marshall sign transformation.\cite{marshall} Upon doping, however, the
fermion signs get active again but it is obvious that the sign structure has
to be quite different from that of a Fermi gas, given that all signs
disappear at half filling.

It is instructive to first specify the sign structure in a Fermi gas. In a
path-integral formalism,\cite{kleinert} the partition function of a Fermi
gas can be expressed as%
\begin{equation}
Z_{\mathrm{FG}}=\sum_{c}(-1)^{N_{\mathrm{ex}}[c]}Z_{0}\left[ c\right]
\label{Zfg}
\end{equation}%
with each path $c$ composed of a set of closed loops of the spatial
trajectories of all fermions and $Z_{0}\left[ c\right] >0$. The sign
structure is then governed by $(-1)^{N_{\mathrm{ex}}[c]}$, with $N_{\mathrm{%
ex}}[c]=N-N_{\mathrm{loop}}[c]$ where $N=\sum_{w}wC_{w}(c)$ is the total
number of fermions and $N_{\mathrm{loop}}[c]=\sum_{w}C_{w}(c)$ the closed
loop number, in which $w$ denotes the number of fermions in a loop (also
called the winding number\cite{kleinert} of the loop) and $C_{w}(c)$, the
number of loops with a given $w$ for a given path $c$.

\begin{figure}[t]
\includegraphics[width=3.5in]{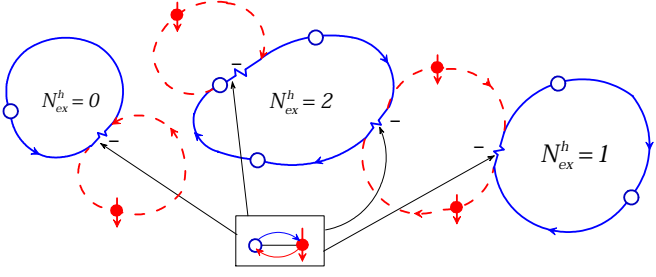}
\caption{A typical diagram for a set of closed paths, denoted by $c$ in the
expansion of the partition function given in (\protect\ref{ztj}). Only the
hole and down spin loops are shown as the up spins are not independent due
to the no double occupancy constraint. Pure spin loops without involving
exchanges with the holes do not contribute to any signs and are not
explicitly shown here. The total sign $\protect\tau _{c}$ associated with
the diagram, defined in (\protect\ref{tauc}), is determined by counting the
hole-down-spin exchanges and the hole permutations. In this particular $c$, $%
N_{h}^{\downarrow }=1+2+1=4$ and $N_{\mathrm{ex}}^{h}=0+2+1=3$ such that $%
\protect\tau _{c}=-1$.}
\label{psf}
\end{figure}

Here we report our discovery of a remarkably sparse sign structure for the
t-J model, which can be rigorously identified at arbitrary doping.
Basically, we shall prove that the partition function for the t-J model is
given by%
\begin{equation}
Z_{\mathrm{t-J}}=\sum_{c}{\tau }_{c}\mathcal{Z}[c]  \label{ztj}
\end{equation}%
where $\mathcal{Z}[c]>0$ [see (\ref{zc})] and the sign structure
\begin{equation}
{\tau }_{c}\equiv (-1)^{N_{h}^{\downarrow }[c]+N_{\mathrm{ex}}^{h}[c]}
\label{tauc}
\end{equation}%
for a given $c$ composed of a set of closed loops for all holes and spins
(an example is shown in Fig. \ref{psf}), where $N_{h}^{\downarrow }[c]$
denotes the total number of exchanges between the holes and down spins and $%
N_{\mathrm{ex}}^{h}[c]$ the total number of exchanges between holes. In
addition to appearing in the above partition function, the sign structure $%
\tau _{c}$ will be also present in various physical quantities based on
expansions in terms of quantum-paths in real space: the $n$-hole propagator
of the Mott-insulating state as well as the zero temperature perturbation
theory of the ground state energy (both up to all orders).

Compared to the full fermion signs in (\ref{Zfg}), which is an exactly
solvable problem for a Fermi gas,\cite{kleinert} \emph{the \textquotedblleft
sign problem\textquotedblright\ for the t-J model then becomes that }$\tau
_{c}$\emph{\ in (\ref{tauc}) is too sparse to be treated as a fermion
perturbative problem.} It implies that in the mathematically equivalent
slave-boson representation, the no double occupancy constraint must play a
crucial role to \textquotedblleft rarefy\textquotedblright\ the statistical
signs of fermionic \textquotedblleft spinons\textquotedblright\ in order to
reproduce the correct sign structure $\tau _{c}$, which disappears at
half-filling. On the other hand, in the slave-fermion representation besides
the statistical signs associated with the fermionic \textquotedblleft
holons\textquotedblright\ [related to $N_{\mathrm{ex}}^{h}$ in ($\ref{tauc})$%
], extra signs in $\tau _{c}$ will have to be generated \emph{dynamically}$,$%
\emph{\ }which are previously known as the phase strings identified in the
one-hole case.\cite{pstring1}

In particular, we will show that in the two-dimensional (2D) case $\tau _{c}$
can be precisely captured by a pair of mutual Chern-Simons gauge fields: the
electrical charges feel $\pi $ flux-tubes attached to the spin
\textquotedblleft particles\textquotedblright\ and vice versa, in an
all-boson formalism which is known as the phase string formulation derived
before by a different method.\cite{pstring2} So the unusual sign structure $%
\tau _{c}$ strongly hints a mutual statistics nature of this doped Mott
insulator, and thus offers critical guidance in the construction of correct
quantum states of it.

Let us begin with the t-J model on a bipartite lattice of any dimensions $%
H_{t-J}=H_{t}+H_{J}$. In the slave-fermion representation, the electron
annihilation operator can be written as%
\begin{equation}
c_{i\sigma }=(-\sigma )^{i}f_{i}^{\dagger }b_{i\sigma }  \label{slave-f}
\end{equation}%
where $f$ denotes the fermionic holon operator and $b$ the bosonic spinon
operator, which satisfy the no double occupancy constraint $f_{i}^{\dagger
}f_{i}+\sum_{\sigma }b_{i\sigma }^{\dagger }b_{i\sigma }=1$. Then the
hopping and superexchange terms can be expressed, respectively, as follows

\begin{equation}
H_{t}=-t(P_{o\uparrow }-P_{o\downarrow })  \label{ht}
\end{equation}%
\begin{equation}
H_{J}=-{\frac{J}{2}}(P_{\uparrow \downarrow }+Q)  \label{hj}
\end{equation}%
where
\begin{eqnarray}
P_{o\uparrow } &=&\sum_{<ij>}f_{i}^{\dagger }f_{j}b_{j\uparrow }^{\dagger
}b_{i\uparrow }+H.c. \\
P_{o\downarrow } &=&\sum_{<ij>}f_{i}^{\dagger }f_{j}b_{j\downarrow
}^{\dagger }b_{i\downarrow }+H.c. \\
P_{\uparrow \downarrow } &=&\sum_{<ij>}b_{i\uparrow }^{\dagger }b_{j\uparrow
}b_{j\downarrow }^{\dagger }b_{i\downarrow }+H.c. \\
Q &=&\sum_{<ij>}\left( n_{i\uparrow }n_{j\downarrow }+n_{i\downarrow
}n_{j\uparrow }\right)  \label{Q}
\end{eqnarray}%
Here $P_{o\uparrow }$ and $P_{o\downarrow }$ denote the nearest neighbor
hole-spin exchange operators, $P_{\uparrow \downarrow }$ the nearest
neighbor spin-spin exchange operator, while the $Q$ term describes the
potential energy between the nearest neighbor antiparallel spins.

Note that the Marshall sign\cite{marshall} factor $(-\sigma )^{i}$ is
explicitly introduced in (\ref{slave-f}) such that the superexchange term $%
H_{J}$ acquires a total negative sign in front of the spin exchange and
potential operators. Then one finds the matrix element $\langle \phi
^{\prime }|H_{J}|\phi \rangle \leq 0$ where $|\phi \rangle $ and $|\phi
^{\prime }\rangle $ denote the Ising spin basis $b_{i_{1}\sigma
_{1}}^{\dagger }b_{i_{2}\sigma _{2}}^{\dagger }...|0\rangle $, which implies
that the $H_{J}$ term will not cause any sign problem. In particular, the
ground state of $H_{J}$ can be always written as
\begin{equation}
|\psi _{0}\rangle =\sum_{\phi }\chi _{\phi }|\phi \rangle \text{ \ \ with \ }%
\chi _{\phi }\geq 0  \label{psi0}
\end{equation}%
which is true even at doped case so long as there is no hopping term.

\emph{Partition function. }The nontrivial sign problem only arises when
holes are doped into the system and allowed to hop. It can be traced to the
sign difference between the hole-spin exchange operators, $P_{o\uparrow }$
and $P_{o\downarrow }$, in the hopping term (\ref{ht}) in addition to the
sign problem associated with fermionic holons. By making the
high-temperature series expansion of the partition function up to all orders
\begin{eqnarray}
Z_{t-J} &=&\text{Tr}\left\{ e^{-\beta H_{t-J}}\right\} =\sum_{n}{\frac{%
(-\beta )^{n}}{{n!}}}\text{Tr}\left\{ (H_{t-J})^{n}\right\}  \notag \\
&=&\sum_{n}{\frac{(+\beta J/2)^{n}}{{n!}}}\text{Tr}\left\{ \sum \cdots
\left( \frac{2t}{J}P_{o\uparrow }\right) \cdots \right.  \notag \\
&&\left. \cdot P_{\downarrow \uparrow }\cdots \left( -\frac{2t}{J}%
P_{o\downarrow }\right) \cdots Q\cdots \right\}  \label{Z}
\end{eqnarray}%
and inserting the complete set
\begin{equation}
\sum_{\phi \{l_{h}\}}|\phi ;\{l_{h}\}\rangle \langle \phi ;\{l_{h}\}|=1
\label{basis}
\end{equation}%
between the operators inside the trace (here $|\phi ;\{l_{h}\}\rangle $ is
an Ising basis with $\phi $ specifying the spin configuration and $\{l_{h}\}$
denoting the positions of holes), one can evaluate term by term of the
expansion in (\ref{Z}). Because of the trace, the initial and final hole and
spin configurations should be the same such that all contributions to $%
Z_{t-J}$ can be characterized by closed loops of holes and spins although
each of them can involve multi-holes or -spins as shown in Fig. \ref{psf}.

Finally one arrives at the compact form given in (\ref{ztj}) based on the
above high-temperature expansion, with

\begin{equation}
\mathcal{Z}[c]=\left( \frac{2t}{J}\right) ^{M_{h}[c]}\sum_{n}{\frac{(\beta
J/2)^{n}}{n!}}\delta _{n,M_{h}+M_{\uparrow \downarrow }+M_{Q}}  \label{zc}
\end{equation}%
in which $M_{h}[c]$ and $M_{\uparrow \downarrow }[c]$ represent the total
steps of the hole and down spin \textquotedblleft
hoppings\textquotedblright\ along the closed loops for a given path $c$, and
$M_{Q}[c]$ the total number of down spins interacting with the
nearest-neighbor up spins via the potential term $Q$ in (\ref{hj}).
Obviously $\mathcal{Z}[c]\geq 0$ in (\ref{zc}). Thus the nontrivial sign
structure of the partition function $Z_{t-J}$ in (\ref{ztj}) is entirely
captured by ${\tau }_{c}$ in (\ref{tauc}) where $N_{h}^{\downarrow }[c]$
denotes the total number of exchanges between the holes and down spins
[i.e., those actions taken via $P_{o\downarrow }$ in (\ref{ht})] and $N_{%
\mathrm{ex}}^{h}[c]$ the total number of exchanges between holes arising
from the fermionic statistics of the holon operator $f$.

The expression (\ref{ztj}) for the partition function clearly demonstrates
that ${\tau }_{c}$ precisely depicts the irreducible sign structure \emph{at
arbitrary doping, temperature, and dimensions }for the t-J model on a
bipartite lattice. In the following, we shall further illustrate how ${\tau }%
_{c}$ similarly appears in other physical quantities.

\emph{Multi-hole propagator. }Define the multi-hole propagator
\begin{eqnarray}
&&G(\{j_{s}\},\{i_{s}\};E)  \notag \\
&=&\langle \psi _{0}|c_{j_{1}\sigma _{1}}^{\dagger }c_{j_{2}\sigma
_{2}}^{\dagger }\cdots G(E)\cdots c_{i_{2}\sigma _{2}}c_{i_{1}\sigma
_{1}}|\psi _{0}\rangle   \label{G}
\end{eqnarray}%
where $|\psi _{0}\rangle $ is the \emph{half-filling} ground state and
\begin{equation}
G(E)=\frac{1}{E-H_{t-J}+0^{+}}.  \label{ge}
\end{equation}%
One can make the following expansion which converges at $E<E_{G}$ (the
multi-hole ground state energy):
\begin{eqnarray}
G(E) &=&{\frac{1}{E}}\sum_{n}\left( \frac{H_{t-J}}{E}\right) ^{n}={\frac{1}{E%
}}\sum_{n}\sum   \notag \\
&&\cdots \left( \frac{t}{-E}P_{o\uparrow }\right) \cdots \left( \frac{J}{-2E}%
P_{\downarrow \uparrow }\right) \cdots   \label{ge-exp}
\end{eqnarray}%
and then insert the complete set (\ref{basis}) between the exchange
operators. Similar to the evaluation of the partition function, denoting $c$
as a given set of \emph{open} paths connecting the hole configurations $%
\{i_{s}\}$ and $\{j_{s}\}$, with $|\psi _{0}\rangle $ expanded in terms of $%
|\phi \rangle $ [(\ref{psi0})], we find
\begin{equation}
G(\{j_{s}\},\{i_{s}\};E)={-}\Lambda \sum_{\phi \phi ^{\prime }}\frac{\chi
_{\phi }\chi _{\phi ^{\prime }}}{-E}\sum_{c}\tau _{c}W[c;E]
\label{propagator}
\end{equation}%
in which each set of paths $c$ is weighed by the phase strings $\tau _{c}$
and an amplitude%
\begin{equation}
W[c;E]=\left( \frac{t}{-E}\right) ^{M_{h}}\left( \frac{J}{-2E}\right)
^{M_{\uparrow \downarrow }+M_{Q}}  \label{W}
\end{equation}%
with $\Lambda =\prod\nolimits_{s=1}^{N_{h}}\left( -\sigma _{s}^{{}}\right)
^{i_{s}-j_{s}}.$ At $E<E_{G}<0$, the expansion (\ref{propagator}) is
converged, and $W[c;E]\geq 0$ shows that $\tau _{c}$ is indeed an
\textquotedblleft irreparable\textquotedblright\ (irreducible) sign which is
expected to play a critical role via constructive and destructive quantum
phase interferences among different \textquotedblleft
path\textquotedblright\ $c$'s. Note that the single-hole version of (\ref{G}%
) has been previously discussed in Ref. \cite{pstring1,pstring2}.

\emph{Ground state wave function.} Define a wave function
\begin{equation}
\Psi _{0}[\mathcal{R}]\equiv \langle \psi _{0}|c_{i_{1}\sigma _{1}}^{\dagger
}c_{i_{2}\sigma _{2}}^{\dagger }\cdots \left\vert \Psi _{0}\right\rangle
\label{gs}
\end{equation}%
with $|\Psi _{0}\rangle $ as the true ground state and $\mathcal{R}\equiv
\{i_{h}\};\{\sigma _{s}\}$. Then, according to (\ref{G}) and (\ref{ge}), $%
\Psi _{0}[\mathcal{R}]$ will be selected as $E\rightarrow E_{G}$ from below,
with (\ref{propagator}) implying

\begin{equation}
\Psi _{0}\left( \mathcal{R}\right) \Psi _{0}^{\ast }\left( \mathcal{R}%
\right) \rightarrow \sum_{c_{\mathcal{R}}}\tau _{c_{\mathcal{R}}}\mathcal{W}%
[c_{\mathcal{R}}]  \label{wf}
\end{equation}%
where on the right hand side the path $c_{R}$'s are all the closed loops
connected to $\mathcal{R}$, each weighed by a positive amplitude $\left.
\mathcal{W}[c_{\mathcal{R}}]=\sum_{\phi \phi ^{\prime }}\chi _{\phi }\chi
_{\phi ^{\prime }}W[c_{\mathcal{R}};E]\frac{E-E_{0}}{E_{0}}\right\vert
_{E\rightarrow E_{0}}$. Therefore the sign structure $\tau _{c_{\mathcal{R}}}
$ must be naturally built into the ground state wave function. In the
following we first examine the ground state energy, and then the mutual
statistics implied for the wave function.

\emph{Ground state energy. }Based on the Goldstone's theorem,\cite{goldstone}
the energy shift of the ground state due to the hopping term $H_{t}$ can be
expressed by

\begin{equation}
E_{G}-\Omega _{0}=\langle \Psi _{0}|H_{t}\sum_{n=0}^{\infty }\left( \frac{1}{%
\Omega _{0}-H_{J}}H_{t}\right) ^{n}|\Psi _{0}\rangle _{\mathrm{connected}}
\label{goldstone}
\end{equation}%
where $H_{J}|\Psi _{0}\rangle =$ $\Omega _{0}|\Psi _{0}\rangle $ and the
subscript \textquotedblleft connected\textquotedblright\ means that only
matrix elements of the operator in (\ref{goldstone}) which start from the
ground state $|\Psi _{0}\rangle $ and end with $|\Psi _{0}\rangle $ without
disconnected parts should be included.

Here $|\Psi _{0}\rangle $ is generally written in a translational invariant
form with a momentum $\mathbf{K}$:%
\begin{equation*}
|\Psi _{0}(\mathbf{K)}\rangle =\sum_{\mathbf{R}}e^{i\mathbf{K}\cdot \mathbf{R%
}}|\Phi _{0};\left\{ \mathbf{r}_{l_{h}}-\mathbf{R}\right\} \rangle
\end{equation*}%
where $|\Phi _{0};\left\{ \mathbf{r}_{l_{h}}\right\} \rangle $ is also the
ground state of $H_{J}$ for a set of hole distribution $\left\{ \mathbf{r}%
_{l_{h}}\right\} $ which minimizes the superexchange energy $\Omega _{0}$.
Like in the half-filling case, one can expand $|\Phi _{0};\left\{ \mathbf{r}%
_{l_{h}}\right\} \rangle $ in terms of the Ising basis: $|\Phi _{0};\left\{
\mathbf{r}_{l_{h}}\right\} \rangle =\sum_{\phi }\chi _{\phi }(\left\{
\mathbf{r}_{l_{h}}\right\} )|\phi ;\left\{ \mathbf{r}_{l_{h}}\right\}
\rangle $ with $\chi _{\phi }\geq 0$ as mentioned before.

By making the expansion in terms of $H_{J}/\Omega _{0}$ and using a similar
procedure in dealing with the expansion (\ref{ge-exp}), one finally gets
\begin{equation}
E_{G}-\Omega _{0}=\Omega _{0}\sum_{\mathbf{RR}^{\prime }}e^{i\mathbf{K}\cdot
(\mathbf{R-R}^{\prime })}\sum_{\phi \phi ^{\prime }}\chi _{\phi ^{\prime
}}\chi _{\phi }\sum_{c(\mathrm{connected)}}^{{}}\tau _{c}W[c;\Omega _{0}]
\label{energy}
\end{equation}%
in which the path $c$ starts from $|\Phi _{0};\left\{ \mathbf{r}_{l_{h}}-%
\mathbf{R}\right\} \rangle $ and ends with $|\Phi _{0};\left\{ \mathbf{r}%
_{l_{h}}-\mathbf{R}^{\prime }\right\} \rangle $ without including the
\textquotedblleft disconnected\textquotedblright\ paths.\cite{goldstone}
Indeed the sign factor $\tau _{c},$ weighed by $W[c;\Omega _{0}]\geq 0$
defined in (\ref{W}), determines the ground state energy shift upon doping.

$\emph{Mutual}$ \emph{statistics. }$\tau _{c}$ in (\ref{tauc}) suggests that
statistics signs of the fermionic holes associated with the slave-fermion
representation are actually indistinguishable from the \textquotedblleft
phase strings\textquotedblright\ generated by the motion of holes. It will
be thus instructive to treat the whole sign structure on an \emph{equal
footing}: take the holon and spinon mathematically as all \emph{bosons} and
redefine the hole-spin and spin-spin exchange operators (\ref{ht}) and (\ref%
{hj}) by
\begin{eqnarray}
P_{o\uparrow } &=&\sum_{\left\langle ij\right\rangle }\left(
e^{-iF_{ij}^{{}}}h_{i}^{\dagger }h_{j}\right) \left( b_{j\uparrow }^{\dagger
}b_{i\uparrow }\right) +H.c.  \label{p1b} \\
-P_{o\downarrow } &=&\sum_{\left\langle ij\right\rangle }\left(
e^{-iF_{ij}^{{}}}h_{i}^{\dagger }h_{j}\right) \left(
e^{-iG_{ji}^{{}}}b_{j\downarrow }^{\dagger }b_{i\downarrow }\right) +H.c.
\label{p2b} \\
P_{\uparrow \downarrow } &=&\sum_{\left\langle ij\right\rangle }\left(
b_{i\uparrow }^{\dagger }b_{j\uparrow }\right) \left(
e^{-iG_{ji}^{{}}}b_{j\downarrow }^{\dagger }b_{i\downarrow }\right) +H.c.
\label{p3b}
\end{eqnarray}%
where the fermionic holon $f_{i}$ is replaced by a \emph{bosonic} $h_{i}$
and the minus sign in front of $P_{o\downarrow }$ is absorbed. The $Q$ term (%
\ref{Q}) remains unchanged. In 2D case, it is straightforward to verify that
if $F_{ij}^{{}}$ and $G_{ij}^{{}}$ are chosen as%
\begin{eqnarray}
F_{ij}^{{}} &=&\sum_{l\neq i,j}[\theta _{i}(l)-\theta
_{j}(l)](n_{l\downarrow }^{b}+n_{l}^{h})  \label{Fij} \\
G_{ij}^{{}} &=&\sum_{l\neq i,j}[\theta _{i}(l)-\theta _{j}(l)]n_{l}^{h}
\label{Gij}
\end{eqnarray}%
where $n_{l\sigma }^{b}$ and $n_{l}^{h}$ are the number operators of spinon (%
$\sigma $) and holon, respectively, and $\theta _{i}(l)=\Im \ln (z_{i}-z_{l})
$ with $z_{i}$ denoting the complex coordinate of site $i$, then the
partition function (\ref{ztj}) can be correctly reproduced. Without $%
F_{ij}^{{}}$ and $G_{ij}^{{}}$, by contrast, one finds ${\tau }_{c}\equiv 1$
in (\ref{ztj}). Namely the sign structure is indeed entirely captured by the
phase factors, $e^{-iF_{ij}^{{}}}$ and $e^{-iG_{ji}^{{}}}$, in this bosonic
formalism.

Rewriting $F_{ij}^{{}}\equiv -$ $A_{ij}^{s}+\phi _{ij}^{0}+A_{ij}^{h}$, and $%
G_{ij}^{{}}\equiv 2A_{ij}^{h}$, and using the constraint $\sum_{\sigma
}n_{l\sigma }^{b}+n_{l}^{h}=1$, one can show that the three link variables, $%
A_{ij}^{s}$, $A_{ij}^{h}$, and $\phi _{ij}^{0}$, satisfy $%
\sum\nolimits_{\Gamma }A_{ij}^{s}=\pm \pi \sum_{l\in \Sigma _{\Gamma
}}\left( n_{l\uparrow }^{b}-n_{l\downarrow }^{b}\right) ,$ and $%
\sum\nolimits_{\Gamma }A_{ij}^{h}=\pm \pi \sum_{l\in \Sigma _{\Gamma
}}n_{l}^{h},$ for a loop $\Gamma $ enclosing an area $\Sigma _{\Gamma }$,
and $\sum\nolimits_{{\large \Box }}\phi _{ij}^{0}=\pm \pi $ for each
plaquette. So they describe $\pi $ flux tubes bound to spinons, holons, and
each plaquette, respectively. Since the Hamiltonian is invariant under gauge
transformations $h_{i}\rightarrow h_{i}e^{i\varphi _{i}},$ $%
A_{ij}^{s}\rightarrow A_{ij}^{s}+(\varphi _{i}-\varphi _{j})$ and $%
b_{i\sigma }\rightarrow b_{i\sigma }e^{i\sigma \theta _{i}},$ $%
A_{ij}^{h}\rightarrow A_{ij}^{h}+(\theta _{i}-\theta _{j})$, the bosonic
holons and spinons carry the gauge charges of $A_{ij}^{s}$ and $A_{ij}^{h}$,
respectively. With the effect of $\tau _{c}$\ described by the mutual
Chern-Simons gauge fields, $A_{ij}^{s}$ and $A_{ij}^{h}$, the t-J model in
the bosonic formalism explicitly becomes a mutual fractional statistics
(mutual semions) problem.\cite{pstring2} Correspondingly, the electron wave
function $\psi _{e}$ of the t-J model can be also expressed in terms of $%
\psi _{b}$ in this bosonic formalism via $\psi _{e}=\mathcal{K}\psi _{b}$,%
\cite{weng05} in which the large gauge transformation $\mathcal{K}$ will
transform by
\begin{equation}
\mathcal{K\longrightarrow }\text{ }\tau _{c}\mathcal{K}  \label{K}
\end{equation}%
under an operation that the hole and spin coordinates are continuously
permuted via a series of nearest neighbor exchanges (with the no double
occupancy obeyed at each step), with the coordinates forming closed loops,
denoted by $c$, after the system back to the original configuration at the
end of the operation. By comparison, the sign structure in (\ref{Zfg}) is
related to the usual antisymmetric fermionic wave functions for a Fermi gas.

In summary, we have demonstrated rigorously that the Hubbard projections
inherent to the physics of doped Mott insulators change the rules of fermion
statistics fundamentally as compared to the Fermi gas. Pending the doping
level, the irreducible sign structure that is of relevance to the physics is
much more sparse in the former and we have shown that at least in real space
expansions these irreducible signs are easy to count. In particular, in the
2D case, we have established a precise relation in which the physical sign
structure of the t-J model is explicitly determined by the mutual
Chern-Simons fields, with the wave function satisfying the mutual statistics.

This does not mean that we have solved the problem -- the \textquotedblleft
mutual Chern-Simons\textquotedblright\ theory\cite{review} of the phase
string formulation is still far from being completely understood. However,
our results open up new alleys for investigation. High temperature
expansions should be revisited to study in detail in what regard the t-J
signs differ from those of a Fermi gas. It would be quite interesting to
find out how the \textquotedblleft irreducible\textquotedblright\ hyper
nodal surfaces of numerically determined $t-J$ model ground states look
like. At the least, it seems possible to critically test Anderson's
conjecture\cite{anderson1} that the ground state of the doped Mott-insulator
has to be orthogonal to that of the Fermi liquid, using the elementary fact
that wave functions having a qualitatively different nodal surface cannot
overlap.

\begin{acknowledgments}
Z.Y.W. acknowledges helpful discussions with S.H. Yang and K.H. Ding. This
work has been supported by the grant no. 10688401 of NSFC.
\end{acknowledgments}

\end{document}